\documentclass[referee]{pasj00}

\SetRunningHead{The revised orbital period change of V617 Sagittarii}{Shi et al. 2013}

\usepackage{times}

\begin{document}

\title{The confirmation and revision on the orbital period
change of the possible type Ia supernova progenitor V617 Sagittarii}
\author{Guang SHI\altaffilmark{1,2,3}, Sheng-Bang QIAN\altaffilmark{1,2,3},
and Eduardo FERN$\mathrm{\acute{A}}$NDEZ LAJ$\mathrm{\acute{U}}$S\altaffilmark{4,5}
}
\altaffiltext{1}{National Astronomical Observatories/Yunnan Observatory,
Chinese Academy of Sciences, P. O. Box 110, 650011 Kunming, China}
\email{shiguang@ynao.ac.cn}
\altaffiltext{2}{Key Laboratory for the Structure and Evolution of Celestial
Objects, Chinese Academy of Sciences}
\altaffiltext{3}{University of the Chinese Academy of Sciences, Beijing 100049, China}
\altaffiltext{4}{Facultad de Ciencias Astr$\mathrm{\acute{o}}$nmicas y Geofsicas,
Universidad Nacional de La Plata Paseo del Bosque s/n, 1900, La Plata,
Pcia. Bs. As., Argentina}
\altaffiltext{5}{Instituto de Astrofisica de La Plata (CCT La plata - CONICET/UNLP), Argentina}


\maketitle

\begin{abstract}
This work reports new photometric results of eclipsing
cataclysmic variable V617 Sagittarii (V617 Sgr).
We analyzed the orbital period change of V617 Sgr, by employing three
new CCD eclipse timings since 2010 along with all the
available data from the literature.
It was found that the orbital period of V617 Sgr undergoes
an obvious long-term increase, which confirms the result revealed by
Steiner et al. (2006).
The rate of orbital period increase was calculated to be ${\dot{P}}$ =
+2.14(0.05) $\times$ 10$^{-7}$ day/year. This suggests the lifetime of the secondary star
will attain to the end in a timescale of 0.97 $\times$ 10$^6$ years
faster than that predicted previously.
In particular, a cyclic variation with a period of 4.5 year
and an amplitude of 2.3 minutes may present
in the O-C diagram. Dominated by the wind-accretion mechanism,
high mass transfer from the low mass secondary to the white dwarf
is expected to sustain in the V Sge-type star V617 Sgr during its long-term evolution.
The mass transfer rate $|\dot{M}_{tr}|$ was estimated to be in the range
of about 2.2 $\times$ 10$^{-7}$ to 5.2 $\times$ 10$^{-7}$ M$_{\odot}$ yr$^{-1}$.
Accordingly, the already massive ($\geq$ 1.2 M$_{\odot}$)
white dwarf primary will process stable nuclear burning,
accrete a fraction of mass from its companion
to reach the standard Chandrasekhar mass limit ($\simeq$ 1.38 M$_{\odot}$),
and ultimately produce a type
Ia supernova (SN Ia) within about 4 $\sim$ 8 $\times$ 10$^{5}$
years or earlier.
\end{abstract}

\section{Introduction}
V617 Sgr was initially identified as a Wolf-Rayet star -- WR 109 in the catalog
of
van der Hucht et al (1981), and also once misapprehended as an irregular variable in the
General Catalogue of Variable Stars (GCVS, Kholopov et al 1987).
On the basis of some photometric and spectroscopic observations
(Steiner et al.
1988, 1999, 2006, 2007; Cieslinski et al. 1999),
it was commonly believed that V617 Sgr is a close eclipsing
cataclysmic variable binary with an orbital period of 0.207 days
(4.98 hr). V617 Sgr was observed to has a median V magnitude of 14.7 mag,
 and its orbital inclination is 72$^{\circ}$.
Moreover, it was detected to
present double eclipse with minima and maxima in the lightcurves,
strong ionized emission lines (such as He II, NV and OVI etc.) in the
optical spectrum, and
high/low photometric states,
which are
quite similar to those seen in V Sge.
So, V617 Sgr was included in a subgroup of several objects,
namely the V Sge-type stars (the others are WX Cen, DI Cru and QU Car, with relatively
low orbital inclinations; Hachisu \& Kato 2003; Oliveira \& Steiner 2004;
Steiner et al. 2006; Kafka et al. 2012). On account of the striking similarities
on the optical spectroscopic and X-ray emission characteristics,
the V Sge-type stars were proposed as the galactic counterpart of Compact Binary Supersoft
X-ray Source (CBSS)
in the Magellanic Clouds (Steiner \& Diaz 1998).

The supersoft X-ray sources are in principle
thought to be the promising candidates of SNe Ia progenitors
(van Teeseling \& King 1998; Knigge et al. 2000; Parthasarathy et al. 2007; Kato 2010),
mostly radiating strong luminosity in the supersoft X-ray spectral range (20 to 80 eV).
This emission is attributed to the hydrostatic nuclear burning on
the surface of a C/O white dwarf.
To make this process happen, a highly mass-accretion rate (about 10$^{-7}$ M$_{\odot}$ yr$^{-1}$)
is demanded, which may originate from two physical channels in systems with the inverted mass ratios.
One is mass transfer from a more massive donor onto
to a less massive white dwarf on a Kelvin-Helmholtz timescale. Then,
the orbital period trends to be decreasing with time (van den Heuvel et al. 1992).
The alternative one causing the high
mass accretion is a very strong wind from the strongly irradiated
low-mass donor (i.e. the wind-accretion scenario),
which should produce an increase in the orbital period
(van Teeseling \& King 1998; Oliveira \& Steiner 2007).

The secular evolution in the eclipsing binary systems can be
investigated
by means of measuring the change of orbital period from the CCD monitoring
of timings with high accuracy. For V617 Sgr with an orbital period of 4.98 hr,
a long-term increase in the orbital period was found,
following the wind-accretion scenario as pointed out by Steiner et al. (2006).
To inspect this result,
new CCD observations on the eclipsing
binary V617 Sgr were implemented during 2010 - 2013 shown in Section 2.
A revised analysis on the orbital period change of V617 Sgr is introduced in Section 3. Finally,
we mainly discuss the observed period evolution and the mass
transfer process that may provide a deep insight
into the nature of the V Sge-type star V617 Sgr
and summary the conclusions in Section 4.

\section{New Observations and Reductions}

\begin{table*}
\caption{The New Observations on the eclipsing binary V617 Sgr\label{tab:1}}
\begin{center}
\begin{tabular}{ccclccl}
  \hline
Obs. Date     &State time(HJD) &End time(HJD) &Eclipse timings(HJD) &Method  &Filter \\
  \hline
			  &(+2400000)       &(+2400000)  &(+2400000)   &      &       \\
 \hline
2010 Jun 10  &55355.80825    &55355.92935  &55355.88437(20) &CCD    &I  \\
2012 Apr 17  &56034.72212    &56034.82578  &56034.782305(135) &CCD    &I  \\
2013 Apr 16  &56398.69203    &56398.82116  &56398.77636(19) &CCD    &I   \\
\hline
\end{tabular}
\end{center}
\end{table*}

The eclipsing binary V617 Sgr was monitored
on 2010 June 8, 2012 April 17 and 2013 April 16, with the 2.15m Jorge Sahade telescope
located at Complejo Astronomico El Leoncito (CASLEO), San Juan, Argentina.
A Roper Scientific Versarray 1300B camera system
with an EEV CCD36-40 de 1340 $\times$ 1300 pix CCD chip was adopted
in the monitoring program. All the CCD
photometric observations were carried out in I passband.
During the observations, the clock of
the control computer was calibrated against UTC time by the GPS receiver's clock.
In the same field of view of the target, we chose two nearby stars with
similar brightness as the comparison star
and the check star, respectively.
All images were corrected after the subtraction of the bias and flat frames,
and reduced by using PHOT (measure magnitudes for a list of stars)
in the aperture photometry package of IRAF.

\begin{figure}
\begin{center}
\FigureFile(80mm,50mm){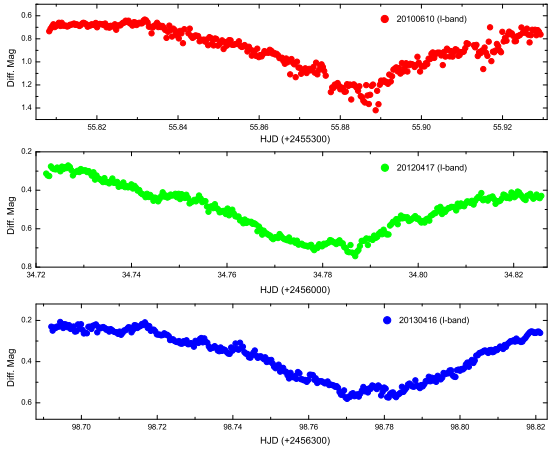}
\caption{The differential light curves of V617 Sgr in the I band measured
on 2010 June 10, 2012 April 17 and 2013 April 16, marked with red,
green and blue points, respectively.\label{fig1}}
\end{center}
\end{figure}

Figure 1 shows the corresponding differential
lightcurves with different eclipse depths.
This may indicate the presence of the high/low states in this source.
The averaged value of eclipse depths is about 0.45 mag.
Fluctuations with timescale of a few tens of minutes and small
flickering can be seen throughout the whole process of eclipse.
During the eclipse minimums on the observations of 2012 Apr 17 and 2013 Apr 16,
the lightcurves are
characterized by brightening with a small bump superimposed
to the flatten-bottom shape
(as shown in the middle and bottom panels of Figure 1).
The width of the flatten bottom shape varies in
two lightcurves.

By means of the least-squares parabolic fitting method,
three new eclipse timings
in Heliocentric Julian Day (HJD) with high time precision
were determined from our CCD photometric data,
which are listed in Table 1.
All available minimum
timings of V617 Sgr since 1987 - 2013
spanning nearly 4.6 $\times$ 10$^4$ orbital cycles
were collected to construct the O-C diagram.
Since the Barycentric Dynamical Time (BJD) is a highly
accurate time system, we converted all the timings
to BJD shown in Table 2.

\section {Analysis of Orbital Period change}

\begin{figure}
\begin{center}
\FigureFile(80mm,50mm){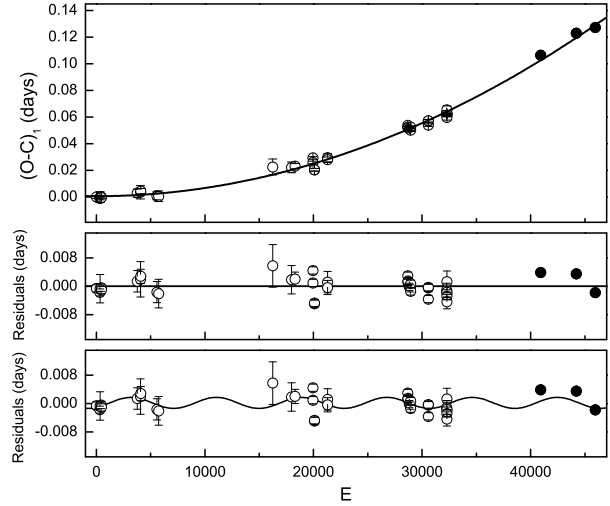}
\caption{Upper panel: the upward parabola fitting curve of the O-C variations
  (open cycles represent the old data from the literature and
  solid cycles represent the new data in this work).
  Middle panel: the residuals of the O-C values.
  Bottom panel: the sine fitting curve for the residuals based
  on the Equation 3.\label{fig2}}
\end{center}
\end{figure}

Some photometric investigations on the orbital period of V617 Sgr
have been performed by several authors
(Steiner 1999, 2006; Cieslinski et al. 1999).
To calculate the O-C values of all 39 eclipse timings,
the obtained linear ephemeris from Steiner et al. (2006)
should be converted to BJD as:
\begin{eqnarray}
\mathrm{Min.I (BJD)} = 2446878.773612 + 0^{d}.20716568 \times E.
\end{eqnarray}
The O-C values with respect to the linear ephemeris are
listed in Table 2.
The revised equation is calculated
by using a least-squares method:
\begin{eqnarray}
\mathrm{Min.I (BJD)}&=&2446878.77421(3)\nonumber\\
&+&0^{d}.20716569(5)\times E\nonumber\\&+&6.07(13) \times 10^{-11} \times E^2,
\end{eqnarray}
with standard deviation $\sigma_1$ = 2$^{d}$.327 $\times$ 10$^{-4}$.
The corresponding fitting curve is displayed in the upper panel of
Figure 2,
while the residuals in the middle panel. The quadratic term
in this
ephemeris is +6.07(13) $\times$ 10$^{-11}$ days/cycle,
which is nearly 10 percents larger than the value +5.5(1) $\times$ 10$^{-11}$
reported in Equation (2) of Steiner et al. (2006).
A revised
positive value of the orbital period variation is $\dot{P}$ = +2.14(0.05)
$\times$ 10$^{-7}$ day/year, leading to an observed timescale of period change
$P/\dot{P}$ = 0.97 $\times$ 10$^{6}$ years.

The residuals related to the parabolic fit still show a small fluctuation,
and may present a cyclical
characteristic. It can be described as:
\begin{eqnarray}
\mathrm{Min.I (BJD)}&=&2446878.77439(12)\nonumber\\&-&0.0013(6)\cos(0.0008 \times E)
\nonumber\\&+&0.0009(6)\sin(0.0008 \times E),
\end{eqnarray}
with standard deviation $\sigma_2$ = 2$^{d}$.023 $\times$ 10$^{-4}$.
The relevant fitting curve is plotted in the bottom panel of Figure 2.
The $\chi^2$ value is $\sim$ 1 far less than the value 9.77 in the parabolic fit.
Based on the F-test proposed by Pringle (1975), F(3,36) = 5.24 are obtained,
which reveals a confidence level above 99.58\% for the sinusoidal ephemeris.
Since the sinusoidal fit is significant, a periodic component added on the
orbital period increase plausibly exist in the O-C variations.
The cyclic variation yields an amplitude of about 2.3 min and a time-scale
of about 4.5 year.

\begin{table*}
\caption{The 39 eclipse timings of the eclipsing binary V617 Sgr\label{tab:2}}
\setlength{\tabcolsep}{2pt}
\begin{center}
\begin{tabular}{cccccccc}
  \hline
BJD            &Errors & E &O-C & Residuals &Ref. \\
  \hline
(+2400000)     & days  &   &days &days &  \\
 \hline
 \noalign{\smallskip}
46878.773612   &0.001    &0      &+0.000000 	&-0.000590  &(1) \\
46947.758615   &0.001    &333    &-0.001109 	&-0.001710  &(1) \\
46952.731615   &0.004    &357    &-0.000081 	&-0.000680  &(1) \\
46973.655616   &0.001    &458    &+0.000205 	&-0.000400  &(1) \\
46974.483616   &0.001    &462    &-0.000457 	&-0.001060  &(1) \\
47658.754647   &0.003    &3765   &+0.002928 	&+0.001430  &(1) \\
47721.733648   &0.005    &4069   &+0.003617 	&+0.001970  &(1) \\
47725.670648   &0.002    &4088   &+0.004472 	&+0.002820  &(1) \\
48036.829660   &0.003    &5590   &+0.000903 	&-0.001640  &(1) \\
48069.768660   &0.004    &5749   &+0.000589 	&-0.002060  &(1) \\
50246.685680   &0.006    &16257  &+0.022535 	&+0.005750  &(1) \\
50602.595690   &0.004    &17975  &+0.022216 	&+0.001840  &(1) \\
50671.582704   &0.002    &18308  &+0.023118 	&+0.002010  &(1) \\
51011.754720   &0.001    &19950  &+0.029383 	&+0.004450  &(1) \\
51013.615720   &0.001    &19959  &+0.025894 	&+0.000940  &(1) \\
51040.541721   &0.001    &20089  &+0.020380 	&-0.004890  &(1) \\
51041.577721   &0.001    &20094  &+0.020552 	&-0.004730  &(1) \\
51290.806745   &0.003    &21297  &+0.029480 	&+0.001160  &(1) \\
51292.669745   &0.002    &21306  &+0.027990 	&-0.000340  &(1) \\
52822.612766   &0.001    &28691  &+0.053794 	&+0.002980  &(2) \\
52823.646766   &0.001    &28696  &+0.051966 	&+0.001140  &(2) \\
52824.682766   &0.001    &28701  &+0.052139 	&+0.001290  &(2) \\
52825.718766   &0.001    &28706  &+0.052311 	&+0.001450  &(2) \\
52849.748765   &0.001    &28822  &+0.051112 	&-0.000150  &(2) \\
52873.572764   &0.001    &28937  &+0.051079 	&-0.000590  &(2) \\
52874.607764   &0.001    &28942  &+0.050251 	&-0.001430  &(2) \\
52875.645764   &0.001    &28947  &+0.052424 	&+0.000710  &(2) \\
53211.669752   &0.001    &30569  &+0.053971 	&-0.003600  &(2) \\
53212.708752   &0.001    &30574  &+0.057143 	&-0.000450  &(2) \\
53213.744752   &0.001    &30579  &+0.057316 	&-0.000290  &(2) \\
53564.688738   &0.002    &32273  &+0.062945 	&-0.001140  &(2) \\
53565.723738   &0.001    &32278  &+0.062117 	&-0.001990  &(2) \\
53566.552738   &0.002    &32282  &+0.062455 	&-0.001660  &(2) \\
53566.758738   &0.001    &32283  &+0.061290 	&-0.002830  &(2) \\
53567.798738   &0.003    &32288  &+0.065462 	&+0.001310  &(2) \\
53568.621738   &0.002    &32292  &+0.059800 	&-0.004360  &(2) \\
55355.885124   &0.00020  &40919  &+0.106418 	&+0.003850  &(3) \\
56034.783081   &0.000135 &44196  &+0.123031 	&+0.003520  &(3) \\
56398.777152   &0.00019  &45953  &+0.127319 	&-0.001810  &(3) \\
\noalign{\smallskip}\hline
\end{tabular}
\end{center}
\centering (1). Steiner et al. 1999; (2). Steiner et al. 2006; (3). this work
\end{table*}

\section{Discussion and Conclusions}
In this work, We present an orbital period analysis of V617 Sge
by adopting three new eclipse
timings together with the data from the literature.
The long-term general trend of its orbital period
shows an evidently continuous increase, confirming the result
of Steiner et al. (2006).
A revised orbital period change is obtained at a rate of
$\dot{P}$ = +2.14(0.05) $\times$ 10$^{-7}$ day/year,
which is larger than that derived before.
The previous O-C analysis only covers about 3.2 $\times$ 10$^4$ cycles,
while that reported here contains all the data form 1987 -- 2013 spanning
about 4.6 $\times$ 10$^4$ orbital cycles.
The positive rate of
the orbital period change deduced from the O-C analysis can
describe a secular variation of the orbital
angular moment in V617 Sgr.
Noted that, an observed timescale of orbital period
variation is estimated to be 0.97 $\times$ 10$^{6}$ year.
Thus, the secondary star of the binary V617 Sgr will accomplish its evolution
faster than the predicted result in Steiner et al. (2006).

The V Sge-type stars are regarded as the counterpart of CBSS in the Galaxy.
The prototype V Sge has a high mass ratio of q = 3.8 and an orbital period of 12.3 hours.
The detected decrease in its orbital period is due to the mass transfer
from the more massive donor to the less massive white dwarf primary
on the Kelvin-Helmholtz time scale.
This can produce a highly mass-accretion rate leading to the supersoft
X-ray radiation via hydrostatic nuclear burning on the surface of the white dwarf.
However, for the systems with period shorter than 6 hours, this mechanism
does not work (Oliveria \& Steiner 2007).
Instead, the mass transfer is driven by the wind accretion produced
by the irradiated low-mass donor (Van Teeseling \& King 1998).
As a member of the V Sge-type stars,
V617 Sgr has an orbital period of 4.98 hours shorter than 6 hours.
The binary system is thought to be composed of a massive
($\geq$ 1.2 M$_{\odot}$) white dwarf primary
and an evolved low-mass (about 0.5 M$_{\odot}$) secondary star (Steiner et al. 2006).
Detected with a long-term increase in the orbital period,
V617 Sgr is particularly in favor of the wind-accretion
evolution scenario.

Supposing that the orbital period increase in V617 Sgr
totally results from the consecutive mass transfer
and/or mass lose of the low-mass secondary star, an experiential
relationship between period change and mass
transfer and/or lose is given by the formula:
\begin{equation}
\frac{\dot{P}}{P} = \frac{3\varsigma-1}{2}\frac{\dot{M}_2}{M_2} \end{equation}
from the Equation (6) of Knigge et al. (2000). Here,
$\varsigma$ is the effective mass-radius index of the
secondary and $\dot{M}_2$ is the mass change rate. We adopt the $\varsigma$ = -1/3
(van Teeseling \& King 1998; Knigge et al. 2000)
for the low-mass secondary of V617 Sgr.
Thus, an averaged mass change rate of the secondary
$\dot{M}_2$ = $\dot{M}_{tr}$ + $\dot{M}_{w2}$
$\simeq$ -5.2 $\times$ 10$^{-7}$ M$_{\odot}$ yr$^{-1}$
is required to produce such a high variation in
the orbital period of V617 Sgr. This also reveals
an upper limit of $|\dot{M}_{tr}|$ = $|\dot{M}_2|$ for the long-term mass transfer from
the secondary to the white dwarf,
when the mass lose of the secondary disappears or can be neglected. Contrarily,
considering the largest wind lose for the irradiated secondary
by the X-ray source (the wind is almost fully ionized),
a maximum rate of wind lose
$\dot{M}_{w2,max}$ $\simeq$ 3.0 $\times$ 10$^{-7}$ M$_{\odot}$ yr$^{-1}$
is derived from the Equation (11) of van Teeseling \& King (1998):
\begin{eqnarray}
\dot{M}_{w2,max} \simeq
-6 \times 10^{-6} \sqrt{M_2 R_2}\;M_{\odot}\;yr^{-1}.
\end{eqnarray}
Here, R$_{2}$ can be calculated to be
R$_{2}$ = R$_{L2}$ $\simeq$ 0.53 R$_{\odot}$,
on the assumption of the Roche-lope geometry for
the secondary star (King \& van Teeseling 1998).
This maximum rate of wind lose is dependent on the basic parameters of the secondary
and tend to a gradual decrease as the evolution of the binary system.
Noted that, it may be overestimated
to some extent because of the possible existence of
an very young disk and the actual variation of wind
lose in different evolutionary states.
Then, we obtain a quite critical lower limit of
$|\dot{M}_{tr}|$ $\simeq$ 2.2 $\times$ 10$^{-7}$
M$_{\odot}$ yr$^{-1}$ for the long-term mass transfer.
Hence, following the wind-accretion channel,
the white dwarf primary in V617 Sgr will experience
stable nuclear burning and accumulate mass efficiently
from the irradiated donor star.
After about 3.5 -- 8.2 $\times$ 10$^5$ years or shorter,
the white dwarf of V617 Sgr will grow mass to the
Chandrasekhar mass limit 1.38 M$_{\odot}$,
exploding as a SN Ia in the Galaxy.

Of particularly interesting, a cyclic component with
an amplitude of 2.3 minutes and a timescale of 4.5 years may exist
in the O-C diagram of V617 Sgr, which is reported in this system for the first time.
To comprehend this characteristic, two kinds of mechanisms are of possibility.
One is the solar-type magnetic activity of the secondary star
acquainted as the Applegate's mechanism (Applegate 1992; Lanza et al. 1998).
In this model, the fractional period change of V617 Sgr $\Delta P/P$
can be calculated to be 6.17 $\times$ 10$^{-6}$.
It is noted that this object perform a behaviour similar to that of
cataclysmic variables above the period gap
in the diagram of period versus amplitude (Baptista et al. 2003).
The other one is the light-travel-time effect
via the presence of a tertiary component.
Until recently, these two explanations causing the
cyclical period change in the O-C analysis
of eclipsing cataclysmic variables are still under discussion
(Baptista et al. 2003; Qian et al. 2007, 2009; Dai et al. 2009, 2010).
However, No one can be totally ruled out.
In the future, more long-term monitoring and more
CCD eclipse timings are in particular expected to
check this preliminary result and to investigate
the nature of V617 Sgr.

This work is supported by Chinese Natural Science Founda-
tion through a key project (No. 11133007). New CCD photo-
metric observations of WX Cen were obtained with the 2.15 m
¡°Jorge Sahade¡± telescope. The authors thank the referee for
useful comments and suggestions that helped to improve the
original manuscript greatly.

\end{document}